\documentclass{article}
\usepackage{amsmath}
\usepackage[psamsfonts]{amssymb}
\usepackage{cmmib57}
\usepackage{diagrams}
\newcommand{\bPf}{\par\vspace*{-4pt}\indent{\sc Proof.}\enskip}
\newcommand{\ePf}{\medskip}
\def\QED{\hskip0.1em\hfill\null\ \null\nobreak\hfill\kern3pt\vbox{\hrule\hbox
   {\vrule\kern1pt\vbox{\kern1.7pt\hbox{$\scriptscriptstyle{QED}$}
    \kern0.2pt}\kern1pt\vrule}\hrule}}

\def\END{\hskip0.1em\hfill\null\ \null\nobreak\hfill\kern3pt\vbox{\hrule\hbox
   {\vrule\kern1pt\vbox{\kern1.7pt\hbox{$\,\,\,\vspace{5pt}$}
    \kern0.2pt}\kern1pt\vrule}\hrule}}
\newtheorem{theorem}{Theorem}
\newtheorem{lemma}{Lemma}
\newtheorem{corollary}{Corollary}
\newtheorem{proposition}{Proposition}
\newtheorem{remark}{Remark}
\newtheorem{definition}{Definition}
\newtheorem{example}{Example}
\newcommand{\bCd}{\bEq\begin{CD}}
\newcommand{\eCd}{\end{CD}\eEq}
\newcommand{\bcd}{\beq\begin{CD}}
\newcommand{\ecd}{\end{CD}\eeq}
\newcommand{\ben}{\begin{enumerate}}
\newcommand{\een}{\end{enumerate}}
\newcommand{\bEq}{\begin{eqnarray}}
\newcommand{\eEq}{\end{eqnarray}}
\newcommand{\beq}{\begin{eqnarray*}}
\newcommand{\eeq}{\end{eqnarray*}}
\newcommand{\bDf}{\begin{definition}\em}
\newcommand{\eDf}{\end{definition}}
\newcommand{\bLm}{\begin{lemma}}
\newcommand{\eLm}{\end{lemma}}
\newcommand{\bPr}{\begin{proposition}}
\newcommand{\ePr}{\end{proposition}}
\newcommand{\bTh}{\begin{theorem}}
\newcommand{\eTh}{\end{theorem}}
\newcommand{\bCr}{\begin{corollary}}
\newcommand{\eCr}{\end{corollary}}
\newcommand{\bRm}{\begin{remark}\em}
\newcommand{\eRm}{\end{remark}}
\newcommand{\bEx}{\begin{example}\em}
\newcommand{\eEx}{\end{example}}
\newcommand{\C}{\mathbb{C}}

\newcommand{\ie}{{\em i.e$.$} }
\newcommand{\eg}{{\em e.g$.$} }
\newcommand{\R}{I\!\!R}

\newcommand{\mto}{\mapsto}

\newcommand{\der}{\partial}

\DeclareMathOperator{\byd}{{\raisebox{.1ex}{:}{=}}}

\newcommand{\ucar}[1]{\underset{#1}{\times}}

\newcommand{\balp}{\boldsymbol{\alp}}

\newcommand{\bmu}{\boldsymbol{\mu}}

\newcommand{\bsig}{\boldsymbol{\sig}}

\newcommand{\cA}{\mathcal{A}}

\newcommand{\cC}{\mathcal{C}}

\newcommand{\cE}{\mathcal{E}}

\newcommand{\cJ}{\mathcal{J}}

\newcommand{\cL}{\mathcal{L}}

\newcommand{\by}{\boldsymbol{y}}

\newcommand{\bA}{\boldsymbol{A}}

\newcommand{\bF}{\boldsymbol{F}}
\newcommand{\bG}{\boldsymbol{G}}

\newcommand{\bP}{\boldsymbol{P}}

\newcommand{\bW}{\boldsymbol{W}}
\newcommand{\bX}{\boldsymbol{X}}
\newcommand{\bY}{\boldsymbol{Y}}

\newcommand{\sub}{\subset}

\newcommand{\wed}{\wedge}
\newcommand{\com}{\!\circ\!}
\newcommand{\ten}{\!\otimes\!}

\newcommand{\alp}{\alpha}

\newcommand{\gam}{\gamma}
\newcommand{\del}{\delta}
\newcommand{\eps}{\epsilon}
\newcommand{\zet}{\zeta}
\newcommand{\tht}{\theta}

\newcommand{\lam}{\lambda}
\newcommand{\sig}{\sigma}

\newcommand{\ome}{\omega}

\newcommand{\Lam}{\Lambda}
\newcommand{\Sig}{\Sigma}
\newcommand{\Ome}{\Omega}

\newcommand{\For}{{\Lambda}}
\newcommand{\Con}{{\mathcal{C}}}
\newcommand{\Hor}{{\mathcal{H}}}
\newcommand{\Var}{{\mathcal{V}}}
\newcommand{\Thd}{{\Theta}}
\title{\large{{\bf Noether identities in Einstein--Dirac theory \\ 
and the Lie derivative of spinor fields
}}}
\author{{\normalsize M.
Palese and E. Winterroth}
\\{\footnotesize Department of Mathematics,
University of Turin}
\\{\footnotesize Via C. Alberto 10, 10123 Turin, Italy}\\ 
{\footnotesize e--mails: 
{\sc marcella.palese@unito.it, ekkehart.winterroth@unito.it}}}
\date{}
\begin{document}

\maketitle

\begin{abstract}
We characterize the Lie derivative of spinor 
fields from a variational point of view by resorting to the theory of the Lie derivative of sections of gauge-natural bundles. Noether 
identities from the gauge-natural invariance of the first  variational derivative of the Einstein(--Cartan)--Dirac Lagrangian provide restrictions on the Lie derivative of fields.

\medskip

\noindent {\bf 2000 MSC}: 58A20,58A32,58E30,58E40.

\noindent {\em Key words}: jet, gauge-natural bundle, Noether identity, spinor field
\end{abstract}

\section{Introduction}

It is nowadays widely recognized the preminent {\it r\^ole} played by the gauge-natural functorial approach
to the geometric description of (classical) field theories~\cite{Ec81,FFFG98,FaFr03,FFP01}. 
Physical fields are assumed to be geometrically represented by sections of fiber bundles 
functorially associated with some jets prolongation of the relevant
principal bundle by means of left actions of Lie groups on manifolds, usually tensor spaces. 
Such an approach enables to functorially define the Lie derivative of 
physical fields with respect to gauge-natural lifts of
(prolongations of) infinitesimal principal automorphisms of the underlying principal
bundle~\cite{KMS93}.

This concept generalizes that of the natural lift of an infinitesimal
automorphism of the basis manifold to the bundle of higher order frames
~\cite{KMS93}. The structure group of the total space is generalized to be the
semidirect product of a differential group on a Lie group
$\bG$ --- the gauge group --- which is not in general a subgroup of a differential group. 
In particular, in the Einstein(-Cartan)-Dirac theory, the coupling between gravitational 
and fermionic fields requires the use
of the concept of a {\it spin-tetrad}, which turns out to be a gauge-natural object (see
\eg Refs.~\cite{FaFr03,GoMa03,GoMa05,Mat03,weinberg72} and references therein).

However, since this construction involves the enlargement of the class of morphisms of the category,
such an approach yields an indeterminacy in the concept of conserved quantities. In fact the vertical 
components of an
infinitesimal principal automorphism are completely independent from the components of its projection on 
the tangent bundle to
the basis manifold.
This implies that there is {\it a priori} no
natural way of relating infinitesimal gauge transfomations with 
infinitesimal transformations of the basis manifold, \eg of 
space-time (see \eg in particular Ref.~\cite{Mat03}). It is generally believed that such an 
indeterminacy is somehow unavoidable, and that {\it ad hoc} restrictions~\cite{Kos66,Lic63} on the
allowed automorphisms of the gauge-natural bundle must be performed, in order to coherently and uniquely define a
geometric concept of the Lie derivative of sections of gauge-natural bundles. For a quite
exaustive review see \eg Refs. \cite{FFR03,GoMa03,GoMa05,Mat03}. 

For reasons which will be clear later, due to its invariance with respect to contact structure induced by jets, we consider
the geometric framework {\it finite order variational sequences}~\cite{Kru90} as the most suitable
for the definition of the class of infinitesimal automorphisms of the
gauge-natural bundle with respect to wich the Lie derivative of fields can be defined unambigously.
In particular, in Refs.~\cite{FPV02,PaWi03,PaWi04,PaWi06,PaWi07} a variational sequence  
on gauge-natural bundles was considered, pointing out some important
properties of the Lie derivative of sections of bundles~\cite{Ec81,KMS93} 
and relative consequences on the content of Noether
Theorems~\cite{Noe18}.  

We stress the very important fact --- although underestimated --- that in 
the category of gauge-natural
bundles it is possible to relate the Lie derivative of sections of bundles with {\it the vertical part} --- {\it not} the
vertical component --- of jet prolongations of gauge-natural lifts of infinitesimal 
principal automorphisms 
(they {\it coincide} up to a sign). 
The concept of the vertical part of a projectable vector
field, together with other important related decompositions 
induced by the contact structure on jet bundles will be defined in
the next Section. 

In Ref.~\cite{PaWi03} for the first time the fact has been pointed out that --- 
when taken as variation vector fields, in
order to derive covariantly and canonically conserved Noether currents --- 
{\it such vertical parts are constrained by
generalized Jacobi equations}. A restriction on the Lie derivative of fields is then 
immediately derived by the simple
request of the covariance of conserved quantities generated by gauge-natural symmetries.
We notice that necessary and sufficient conditions (Bergmann-Bianchi identities~\cite{AnBe51}) 
for the existence of canonical
covariant conserved currents and associated superpotentials can be suitably 
interpreted as Noether identities~\cite{Winterroth07}.
By representing the Noether Theorems~\cite{Noe18} in terms of the
generalized gauge-natural Jacobi morphism, the Lie derivative of spinor 
fields is then accordingly characterized as the
above mentioned indeterminacy disappears along the kernel
of the generalized gauge-natural Jacobi morphism.

\section{Variational sequences on jets of gauge-natural bundles}

Consider a fibered manifold $\pi : \bY \to \bX$,
with $\dim \bX = n$ and $\dim \bY = n+m$.
For $s \geq q \geq 0$ integers we are concerned with the $s$--jet space $J_s\bY$ of 
$s$--jet prolongations of (local) sections
of $\pi$; in particular, we set $J_0\bY \equiv \bY$. We recall the natural fiberings
$\pi^s_q: J_s\bY \to J_q\bY$, $s \geq q$, $\pi^s: J_s\bY \to \bX$, and,
among these, the {\it affine} fiberings $\pi^{s}_{s-1}$~\cite{Sau89}.

By adopting a multiindex notation, the charts induced on $J_s\bY$ are denoted by
$(x^\sig,y^i_{\balp})$, with
$0 \leq |\balp| \leq s$; in particular, we set $y^i_{\bf{0}}
\equiv y^i$. The local vector fields and forms of $J_s\bY$ induced by
the above coordinates are denoted by $(\der^{\balp}_i)$ and $(d^i_{\balp})$,
respectively.

Let $\cC^{*}_{s-1}[\bY] \simeq J_s\bY \ucar{J_{s-1}\bY} V^*J_{s-1}\bY$. 
For $s\geq 1$, we have a natural splitting:
\bEq
\label{jet connection}
J_{s}\bY\ucar{J_{s-1}\bY}T^*J_{s-1}\bY =\left(
J_s\bY\ucar{J_{s-1}\bY}T^*\bX\right) \oplus\cC^{*}_{s-1}[\bY]\,.
\eEq

Given a vector field $\Xi : J_{s}\bY \to TJ_{s}\bY$, the splitting
\eqref{jet connection} yields $\Xi \, \com \, \pi^{s+1}_{s} = \Xi_{H} + \Xi_{V}$
where, if $\Xi = \Xi^{\gam}\der_{\gam} + \Xi^i_{\balp}\der^{\balp}_i$, then we
have $\Xi_{H} = \Xi^{\gam}D_{\gam}$ and the invariant expression
$\Xi_{V} = (\Xi^i_{\balp} - y^i_{\balp + \gam}\Xi^{\gam}) 
\der^{\balp}_{i}$. We shall call $\Xi_{H}$ and $\Xi_{V}$ the 
horizontal and {\it the vertical part of $\Xi$}, respectively.
The splitting
\eqref{jet connection} induces also a decomposition of the
exterior differential on $\bY$,
$(\pi^{s}_{s-1})^*\com \,d = d_H + d_V$, where $d_H$ and $d_V$, the {\it horizontal} and {\it
vertical differential}~\cite{FPV98a,Sau89,Vit98}.

Let $\bP\to\bX$ be a principal bundle with structure group $\bG$.
Let $r\leq k$ be integers and $\bW^{(r,k)}\bP$ be the gauge-natural prolongation of $\bP$
with structure group
$\bW^{(r,k)}_{n}\bG$~\cite{Ec81,KMS93}.
Let $\bF$ be any manifold and $\zet: \bW^{(r,k)}_{n}\bG\ucar{}\bF\to\bF$ be 
a left action of $\bW^{(r,k)}_{n}\bG$ on $\bF$. There is a naturally defined 
right action of $\bW^{(r,k)}_{n}\bG$ on $\bW^{(r,k)}\bP \times \bF$ 
so that we can associate in a standard way
to $\bW^{(r,k)}\bP$ the {\it gauge-natural bundle} of order 
$(r,k)$ $\bY_{\zet} \doteq \bW^{(r,k)}\bP\times_{\zet}\bF$~\cite{Ec81,KMS93}. 

Let $\cA^{(r,k)}\doteq T\bW^{(r,k)}\bP/\bW^{(r,k)}\bG$ ($r\leq k$)
be the {\it vector} bundle  over $\bX$ of right invariant 
infinitesimal automorphisms of $\bW^{(r,k)}\bP$. 
The {\it gauge-natural lift} $\mathfrak{G}$ functorially associates with any right-invariant local
automorphism
$(\Phi,\phi)$ of the  principal bundle $W^{(r,k)}\bP$ a unique local automorphism 
$(\Phi_{\zet},\phi)$ of the associated bundle $\bY_{\zet}$. 
An infinitesimal version can be defined:
\bEq
\mathfrak{G} : \bY_{\zet} \ucar{\bX} \cA^{(r,k)} \to T\bY_{\zet} \,:
(\by,\bar{\Xi}) \mto \hat{\Xi} (\by) \,,
\eEq
where, for any $\by \in \bY_{\zet}$, one sets: $\hat{\Xi}(\by)=
\frac{d}{dt} [(\Phi_{\zet \,t})(\by)]_{t=0}$,
and $\Phi_{\zet \,t}$ denotes the (local) flow corresponding to the 
gauge-natural lift of $\Phi_{t}$.
This mapping fulfils important linearity properties~\cite{FFP01}. 

Although the jet prolongation
up to a given order of a gauge-natural bundle is again a gauge-natural bundle associated with some gauge
natural prolongation of the underlying principal bundle~\cite{KMS93},  in general --- as it can be easily seen also from the corresponding
 local (invariant) expression ---  the generalized
vector field $j_{s}\Xi_{V}\doteq (j_{s}\Xi)_{V}$  {\it is not the gauge-natural lift of some infinitesimal principal automorphism}. 

Following Ref.~\cite{KMS93} we give the definition of the Lie derivative of a section of a gauge-natural bundle. Notice that this object is uniquely and functorially defined by the right invariant vector field $\bar{\Xi}$.
\bDf
Let $\gam$ be a (local) section of $\bY_{\zet}$, $\bar{\Xi}$ 
$\in \cA^{(r,k)}$ and $\hat\Xi$ its gauge-natural lift. 
We
define the {\it 
generalized Lie derivative} of $\gam$ along the vector field 
$\hat{\Xi}$ to be the (local) section $\pounds_{\bar{\Xi}} \gam : \bX \to V\bY_{\zet}$, 
given by
$\pounds_{\bar{\Xi}} \gam = T\gam \circ \xi - \hat{\Xi} \circ \gam$.\END
\eDf

\bRm\label{lie}
The Lie derivative of sections
is an homomorphism of Lie algebras; furthermore
for any local section $\gam$ of $\bY_{\zet}$, the mapping 
$\bar{\Xi} \mto \pounds_{\bar{\Xi}}\gam$ 
is a linear differential operator.
As a consequence, for any gauge-natural lift
 $\hat{\Xi}$,
the fundamental relation hold true:
\bEq\label{oh!oh!}
\hat{\Xi}_V= - \pounds_{\bar{\Xi}}\,.\END
\eEq
\eRm

\subsection{Variational derivatives and Noether identities}

Let us now introduce variational sequences on gauge-natural bundles.
For $s \geq 0$ ({\it resp.} $0 \leq q \leq s$) we consider the sheaves $\For^{p}_{s}$
of $p$--forms on $J_s\bY_{\zet}$ ({\it resp.} $\Hor^{p}_{(s,q)}$ and
$\Hor^{p}_{s}$ of {\it horizontal forms} with respect to the 
projections $\pi^s_q$ and $\pi^s_0$).
Furthermore, for $0 \leq q < s$, let $\Con^{p}_{(s,q)}
\sub \Hor^{p}_{(s,q)}$ and $\Con^{p}{_s} \sub
\Con^{p}_{(s+1,s)}$ be the sheaves of {\it contact forms}, \ie horizontal forms 
valued into $\cC^{*}_{s}[\bY_{\zet}]$.
The fibered splitting
\eqref{jet connection} yields the {\em sheaf splitting}
$\Hor^{p}_{(s+1,s)}$ $=$ $\bigoplus_{t=0}^p$
$\Con^{p-t}_{(s+1,s)}$ $\wed\Hor^{t}_{s+1}$. Let the surjective map
$h$ be the restriction to $\For^{p}_{s}$ of the projection of
the above splitting onto the non--trivial summand with the highest
value of $t$. Set $\Thd^{*}_{s}$ $\byd$ $\ker h$ $+$
$d\ker h$~\cite{Kru90,Vit98}.

The following {\em $s$--th order
variational sequence} associated with the fibered manifold
$\bY_{\zet}\to\bX$, where the integer $I$ depends on the dimension of the fibers of $\bY_{\zet}$, is an exact resolution of the constant sheaf $\R_{\bY_{\zet}}$ over $\bY_{\zet}$~\cite{Kru90}:
\beq
\diagramstyle[size=0.1em]
\begin{diagram}
0 & \rTo & \R_{\bY_{\zet}} & \rTo & \For^{0}_s & \rTo^{\cE_{0}} &
\For^{1}_s/\Thd^{1}_s & \rTo^{\cE_{1}} & 
\dots & \rTo^{\cE_{I-1}} & \For^{I}_s/\Thd^{I}_s & \rTo^{\cE_{I}} &
\For^{I+1}_s & \rTo^{d} & 0 \, .
\end{diagram}
\eeq

For our purposes we 
refer to the representation of a truncated variational sequence due to Vitolo~\cite{Vit98}
where $\For^{p}_s/\Thd^{p}_s \equiv 
\Con^{p-n}_{s}\wed\Hor^{n,}{_{s+1}^h}/h(d\ker h)$ with $0\leq p\leq n+2$. 
Further developments can be found \eg in
Refs.~\cite{FPV98a,FPV02,KrMu03,PaWi03,PaWi04,PaWi06,PaWi07,Pom95,Vit98}.

Let now
$\eta\in\Con^{1}_{s}\wed\Con^{1}_{(s,0)}\wed\Hor^{n,}{_{s+1}^{h}}$;
then there is a unique morphism~\cite{KoVi03,Vit98}
$
K_{\eta} \in \Con^{1}_{(2s,s)}\otimes\Con^{1}_{(2s,0)}\wed\Hor^{n,}{_{2s+1}^{h}}
$
such that, for all $\Xi:\bY_{\zet}\to V\bY_{\zet}$,
$C^{1}_{1} (j_{2s}\Xi\ten K_{\eta})=E_{{j_{s}\Xi}\rfloor \eta}$, where $E$ 
is the the generalized Euler--Lagrange 
form~\cite{Kru90}; $C^1_1$ stands for tensor
contraction on the first factor and $\rfloor$ denotes inner product. 

Let $\cL_{j_{s}\Xi}$ be the {\it variational Lie derivative}~\cite{FPV98a}. 
The First and the Second Noether Theorem~\cite{Noe18} then read as follows 
(compare with Ref.~\cite{Tra67}).

\bTh\label{noether I}
Let $[\alp] = h(\alp)$ $\in$ $\Var^{n}_{s}$. Then we
have {\it locally}
\beq
\cL_{j_{s}\Xi}(h(\alp)) =
\Xi_{V} \rfloor \cE_{n}(h(\alp))+
d_{H}(j_{2s}\Xi_{V} \rfloor p_{d_{V}h(\alp)}+ \xi \rfloor h(\alp))\,,
\eeq
where $p_{d_{V}h(\alp)}$ is the generalized momentum associated with $h(\alp)$~\cite{FPV98a,Vit98}.
\eTh

\bTh\label{GeneralJacobi}
Let $\alp\in\For^{n+1}_{s}$. Then we have {\it globally}
\beq 
\cL_{j_{s}\Xi} [\alp] =
\cE_{n}({j_{s+1}\Xi_{V} \rfloor h(\alp)}) +
C^{1}_{1}(j_{s}\Xi_{V}\ten K_{hd\alp}) \,.
\eeq
\eTh

\bDf\label{gn}
We say $\lam$ to be a
{\it gauge-natural invariant Lagrangian} if the gauge-natural lift
$(\hat{\Xi},\xi)$ of {\it any} vector
field $\bar{\Xi} \in \cA^{(r,k)}$ is a  symmetry for
$\lam$, \ie if $\cL_{j_{s+1}\bar{\Xi}}\,\lam = 0$.
In this case the projectable vector field
$\hat{\Xi}$ is
called a {\it gauge-natural symmetry} of $\lam$.\END
\eDf

Let $\lam$ be a Lagrangian and let $\hat{\Xi}_{V}$ be considered a variation vector field. Let us
set $\chi(\lam,\hat{\Xi}_{V})$ $\doteq$ 
 $C^{1}_{1} (\hat{\Xi}_{V}$ $\ten$ $K_{hd\cL_{j_{2s}\bar{\Xi}_{V}}\lam})$ $\equiv$  
$E_{j_{s}\hat{\Xi}\rfloor
hd\cL_{j_{2s+1}\bar{\Xi}_V}\lam}$. Because of linearity properties of
$K_{hd\cL_{j_{2s}\bar{\Xi}_{V}}\lam}$~\cite{KoVi03}, by using a global decomposition formula
due to Kol\'a\v{r}~\cite{Kol83}, we can decompose the morphism defined above as 
$\chi(\lam,\hat{\Xi}_{V})=E_{\chi(\lam,\hat{\Xi}_{V})}+ F_{\chi(\lam,\hat{\Xi}_{V})}$, where $
F_{\chi(\lam,\hat{\Xi}_{V})}$ is a {\it local} horizontal differential which can be globalized by
means of the fixing of a connection; however we will not fix any connection {\it a priori} in the
present paper. Such a decomposition is a kind of integration by parts, which provides us with a
globally defined gauge-natural morphism playing a 
relevant {\it r\^ole}~\cite{PaWi03}. 
\bDf
We call the morphism $\cJ(\lam,\hat{\Xi}_{V})$ $\doteq$
$E_{\chi(\lam,\hat{\Xi}_{V})}$  the {\it gauge-natural generalized Jacobi 
morphism} associated with the Lagrangian $\lam$ and the variation vector field
$\hat{\Xi}_{V}$. 
\END\eDf
Such a morphisms has been also represented on finite order variational sequence modulo
horizontal differentials~\cite{FPV02} and thereby proved to be self-adjont along solutions of the
Euler--Lagrange equations, a result already well known for first order field theories~\cite{GoSt73}. 
The same property has been
also proved in finite order variational sequences 
on gauge-natural bundles~\cite{PaWi07} {\it without quotienting out horizontal differentials}.

Because of linearity properties of the Lie derivative of sections of gauge-natural bundles, 
we can consider the 
form $\ome(\lam,\hat{\Xi}_{V})\doteq -\pounds_{\bar{\Xi}} 
\rfloor \cE_{n} (\lam)$ as a new Lagrangian defined on an extended space 
$J_{2s}(\cA^{(r,k)}\times_{\bY} \bY)$. This Lagrangian plays a very important
{\it r\^ole} in the study of conserved quantities. In fact, it is for example
remarkable that when $\ome(\lam,\hat{\Xi}_{V})$ is an horizontal differential
(\ie a null Lagrangian) from the First Noether Theorem \ref{noether I} we get a
conservation law which holds true along any section of the gauge natural bundle
(not only along solutions of the Euler--Lagrange equations).

It is also remarkable that the new Lagrangian $\ome$, in principle, {\it is not gauge-natural invariant}. 
In fact from the gauge-natural invariance of $\lam$ we only infer  that, for any $\hat{\Xi}$, 
$\cL_{j_{s+1}\hat{\Xi}}[\cL_{j_{s+1}\hat{\Xi}_{V}}\lam]=
\cL_{j_{s+1}[\hat{\Xi},\hat{\Xi}_{V}]}\lam+
\cL_{j_{s+1}\hat{\Xi}_{V}}\cL_{j_{s+1}\hat{\Xi}}\lam=
\cL_{j_{s+1}[\hat{\Xi}_{H},\hat{\Xi}_{V}]}\lam$ and {\it a priori} neither
$[\hat{\Xi}_{H},\hat{\Xi}_{V}] = 0$ {\it nor}
 it is the
gauge-natural lift of some infinitesimal principal automorphism. Nevertheless
it is still possible to derive some  invariance properties of the new Lagrangian
$\ome(\lam,\hat{\Xi}_{V})$ restricted along the kernel of the gauge-natural
generalized gauge-natural Jacobi morphism as well as corresponding Noether conservation laws 
and Noether identities~\cite{Noe18}.
 
Let then $\del^{2}_{\mathfrak{G}}\lam \doteq
\cL_{j_{s+1}\hat{\Xi}_V}\cL_{j_{s+1}\hat{\Xi}_V}\lam$ be the {\it gauge-natural second
variation} of $\lam$ taken  with respect to vertical parts of gauge-natural 
lifts of infinitesimal 
principal automorphisms.
First we generalize a classical result~\cite{GoSt73}.

\bPr
Let $\hat{\Xi}_{V}\in
\mathfrak{K}$, where $\mathfrak{K}$ is the kernel of the gauge-natural Jacobi morphism. We have:
\beq
\cL_{j_{s+1}\hat{\Xi}_{H}}[\cL_{j_{s+1}\hat{\Xi}_{V}}\lam]\equiv 0\,.
\eeq
\ePr

\bPf
\beq
\cL_{j_{s+1}\hat{\Xi}}[\cL_{j_{s+1}\hat{\Xi}_{V}}\lam]=\cL_{j_{s+1}
\hat{\Xi}_{H}}[\cL_{j_{s+1}\hat{\Xi}_{V}} 
\lam ]+\\ +\cL_{j_{s+1}\hat{\Xi}_{V}}[\cL_{j_{s+1}\hat{\Xi}_{V}}\lam ]=
\cL_{j_{s+1}\hat{\Xi}_{H}}[\cL_{j_{s+1}\hat{\Xi}_{V}}\lam]+ \del^{2}_{\mathfrak{G}}\lam\,.
\eeq
Theorem \ref{GeneralJacobi} implies 
$\del^{2}_{\mathfrak{G}}\lam  = \cJ(\lam,\hat{\Xi}_{V})$, thus we get the assertion.
\QED\ePf

\bTh
The existence of canonical covariant conserved currents and superpotentials associated with a gauge-natural invariant Lagrangian is equivalent to the existence of Noether identities associated with the invariance properties of the first  variational derivative of the Lagrangian $\lam$ taken with respect to vertical parts of gauge-natural lifts lying in $\mathfrak{K}$.
\eTh
\bPf
From the above Proposition we see that $\cL_{j_{s+1}\hat{\Xi}}[\cL_{j_{s+1}\hat{\Xi}_{V}}\lam]\equiv 0$. 
This last condition means that $\bar{\Xi}$ is a gauge-natural symmetry of the new Lagrangian
$\cL_{j_{s+1}\hat{\Xi}_{V}}\lam=\ome(\lam,\mathfrak{K})$. This invariance implies also the existence of  Noether identities from the gauge-natural invariance of the Euler--Lagrange morphism
$\cE_{n}(\hat{\Xi}_{V}\rfloor \cE_{n}(\lam))$ and ultimately from the corresponding  invariance properties of the first  variational derivative of the Lagrangian $\lam$. It is easy to verify 
that Bergmann-Bianchi 
identities for the existence of  canonical covariant conserved currents and superpotentials associated with the invariance of $\lam$ coincide with the condition $\hat{\Xi}_{V}\rfloor\cE_{n}(\hat{\Xi}_{V}\rfloor \cE_{n}(\lam))=0$ (see \eg Refs.~\cite{PaWi03,PaWi07} and~\cite{FPV02,GoSt73}) equivalent to $\cL_{j_{s+1}\hat{\Xi}_{V}}\cE_{n}(\hat{\Xi}_{V}\rfloor \cE_{n}(\lam))=0$.
\QED
\ePf

\section{Einstein(--Cartan)--Dirac theory}

Let $\bX$ be a $4$-dimensional manifold admitting Lorentzian structures 
($\textstyle{SO}(1,3)^{e}$-reductions) 
and let $\Lam$ be the epimorphism wich exhibits $\textstyle{SPIN}(1,3)^{e}$ 
as the twofold covering of
$\textstyle{SO}(1,3)^{e}$~\cite{FFFG98,FaFr03}.

We recall that a free spin structure on $\bX$ is a $\textstyle{SPIN}(1,3)^{e}$-principal bundle 
$\pi: \Sig \to \bX$ and a bundle map inducing 
a spin-frame on $\Sig$ given by $\tilde{\Lam}: \Sig\to L(\bX)$ defining a
metric  $g$ {\it via} the reduced subbundle  $\textstyle{SO}(\bX,g)=\tilde{\Lam}(\Sig)$ of $L(\bX)$~\cite{FaFr03,GoMa05,Mat03}.

Now, let $\rho$ be the left action of the group $W^{(1,0)} \textstyle{SPIN}(1,3)^{e}$ on 
the manifold $GL(4,\R)$ given by $\rho: ((A,S),\tht)\mto \Lam(S)\circ \tht\circ A^{-1}$ and 
consider the associated
bundle (a gauge-natural bundle of order $(1,0)$)
\beq
\Sig_{\rho}\doteq W^{(1,0)} \Sig\times_{\rho} GL(4,\R)\,:  \eeq
the {\it bundle of spin-tetrads $\tht$}~\cite{weinberg72}.

The induced metric is $g_{\mu\nu}=\tht^{a}_{\mu}\tht^{b}_{\nu}\eta_{ab}$, where $\tht^{a}_{\mu}$ are 
local components of a spin tetrad $\tht$ and $\eta_{ab}$ the Minkowski metric.

Let $\mathfrak{so}(1,3)\simeq \mathfrak{spin}(1,3)$ be the Lie algebra of 
$\textstyle{SO}(1,3)$. One can consider the left action of $W^{(1,1)} \textstyle{SPIN}(1,3)^{e}$ on
the vector space $(\R^{4})^{*}\ten \mathfrak{so}(1,3)$.

The associated bundle is a gauge-natural bundle of order $(1,1)$:
\beq
\Sig_{l}\doteq W^{(1,1)} \Sig\times_{l}((\R^{4})^{*}\ten \mathfrak{so}(1,3))
\simeq J_{1}(\Sig/Z_{2})/\textstyle{SO}(1,3)^{e}\,,
\eeq
the {\it bundle of spin-connections $\ome$}.

If $\hat{\gam}$ is the linear representation of $\textstyle{SPIN}(1,3)^{e}$ 
on the vector space $\C^{4}$ induced by the choice of matrices $\gam$ we get a $(0,0)$-gauge-natural
bundle
\beq
\Sig_{\hat{\gam}}\doteq \Sig\times_{\hat{\gam}}\C^{4}\,,
\eeq
the bundle of spinors.
A spinor connection $\tilde{\ome}=(\textstyle{id}\ten (T_{e}\Lam)^{-1})(\ome)$ - 
where $T_{e}\Lam$ defines the  isomorphism of Lie algebras $\mathfrak{spin}(1,3) \simeq
\mathfrak{so}(1,3)$ -  is locally given by $\tilde{\ome}_{\mu}=-\frac{1}{4}\ome^{ab}_{\mu}\gam_{ab}$.

In the following the Einstein--Cartan Lagrangian will be the base preserving morphism
$\lam_{EC}: \Sig_{\rho}\ucar{\bX}J_{1}\Sig_{l}\to \Lam^{4}T^{*}\bX$,
locally given by $\lam_{EC}(\tht,J^{1}\ome)\doteq -\frac{1}{2k}\Ome_{ab}\wed\eps^{ab}$, where 
$\Ome_{ab}$ is the curvature of $\ome$, $k=\frac{8\pi G}{c^{4}}$, $\eps^{ab}=e_{a}\rfloor(e_{b}\rfloor
\eps)$ and $\eps$ is the standard volum form on $\bX$. Locally $\eps=\textstyle{det}||\tht
||d_{0}\wed\dots d_{3}$, $e_{a}\doteq e_{a}^{\mu}\der_{\mu}$, where $|| e_{a}^{\mu} ||$ is the inverse
of $||\tht ||$.

The Dirac Lagrangian is the base preserving morphism \\
$\lam_{D}: \Sig_{\rho}\ucar{\bX}\Sig_{l} \ucar{\bX}J_{1}\Sig_{\hat{\gam}}\to \Lam^{4}T^{*}\bX$,
locally given by $\lam_{D}(\tht,\ome,J^{1}\psi )\doteq (\frac{i\alp}{2}
(\bar{\psi}\gam^{a}\nabla_{a}\psi-\nabla_{a}\bar{\psi}\gam^{a}\psi) -m\bar{\psi}\psi)\eps$, whit
$\alp=\/h c$, $\nabla$ the covariant derivative with respect to a connection on $\Sig_{\rho}$, and
$\bar{\psi}$ the adjoint with respect to the standard $\textstyle{SPIN}(1,3)^{e}$-invariant product on
$\C^{4}$. Under the assumption of a minimal coupling the total Lagrangian of a gravitational field
interacting with spinor matter is
$\lam=\lam_{EC}+\lam_{D}$.

Let now $\bar{\Xi}$ be a $\textstyle{SPIN}(1,3)^{e}$-invariant vector field on $\Sig$. 
The lagrangian $\lam$ is invariant with respect to any lift $\hat{\Xi}$ of $\bar{\Xi}$ to the total
space of the theory. 
By the First Noether Theorem the folloving conserved  
Noether current has been found in Ref.~\cite{GoMaVi01}:
\beq
\eps(\lam, \bar{\Xi})= \xi^{b}(-\frac{1}{k}G^{a}_{b}+T^{a}_{b})\eps_{a}   
+\bar{\Xi}^{a\,b}_{v} (\frac{1}{2k}\mathfrak{D}\eps_{ab} -
u_{ab}^{c}\eps_{c})+d_{H}(\frac{1}{2k}\bar{\Xi}^{a\,b}_{v}\eps_{ab})\,,
\eeq 
where $\pounds_{\bar{\Xi}}\ome^{a}_{b}=\xi\rfloor\Ome^{a}_{b} +\mathfrak{D} 
\bar{\Xi}^{a}_{v\,b}$, $\mathfrak{D}$ is the covariant exterior derivative with respect to the
connection $\ome$, and $\bar{\Xi}^{a}_{v\,b}=\bar{\Xi}^{a}_{b} -\ome^{a}_{b\mu}\xi^{\mu}$ is the
vertical part of $\bar{\Xi}$ with respect to $\ome$.
The corresponding superpotential is 
$\nu(\lam,\bar{\Xi})\doteq -\frac{1}{2k}\bar{\Xi}_{v}^{ab}\eps_{ab}$. 

 \subsection{Lie derivative of spinor fields}

The natural splitting induced by the contact structure provides us with the vertical part 
$\hat{\Xi}_{V}^{\bA}= \bar{\Xi}^{\bA} -u^{\bA}_{\mu}\xi^{\mu}$,
where $(x^{\mu},u^{\bA})$ are local coordinates on the total gauge-natural bundle and $\bA$ is a
multiindex. On the other hand , since the vertical part with respect to the spinor-connection
$\tilde{\ome}$ is given by
$\hat{\Xi}_{v}^{\bA}= \hat{\Xi}^{\bA} -\tilde{\ome}^{\bA}_{\mu}\xi^{\mu}$,
we get the simple invariant relation:
\bEq\label{oh!}
\hat{\Xi}_{V}^{\bA}=\hat{\Xi}_{v}^{\bA} +
\tilde{\ome}^{\bA}_{\mu}\xi^{\mu}-u^{\bA}_{\mu}\xi^{\mu}\,.
\eEq

In the above Section we saw that, without the fixing of a connection {\it a priori}, the existence
of {\it canonical} global  conserved quantities in
field theory is related with gauge-natural invariant properties of 
$-\pounds_{\bar{\Xi}} \rfloor \cE_{n} (\lam)$ and
corresponding Noether identities:
\bEq\label{liederspin}
(-1)^{|\bsig|}D_{\bsig}\,(D_{\bmu}
(- \pounds_{\bar{\Xi}}\psi)^{ab}(\der_{cd}(\der^{\bmu}_{ab}(-\frac{1}{2k}\Ome_{ab}\wed\eps^{ab}+ 
\nonumber \\ \nonumber +(\frac{i\alp}{2}(\bar{\psi}\gam^{a}\nabla_{a}\psi-\nabla_{a}\bar{\psi}
\gam^{a}\psi) 
- m\bar{\psi}\psi)\eps)) - \\ 
-\sum_{|\balp |= 0}^{s-|\bmu |}
(-1)^{|\bmu +\balp |} \frac{(\bmu +
\balp)!}{\bmu ! \balp !}
D_{\balp}\der^{\balp}_{cd}(\der^{\bmu}_{ab}(-\frac{1}{2k}\Ome_{ab}\wed\eps^{ab}+ \\ 
\nonumber +(\frac{i\alp}{2}(\bar{\psi}\gam^{a}\nabla_{a}\psi-\nabla_{a}\bar{\psi}\gam^{a}\psi) 
-m\bar{\psi}\psi)\eps))))=0\,.,
\nonumber
\eEq
with $0\leq |\bsig|\leq 1$, $0\leq |\bmu|\leq 1$ and the fibered local coordinates in the total
bundle we denote by $(x^{\mu},
\tht^{a}_{\mu},\ome^{ab},\ome^{ab}_{\mu},\psi)$.
Such identities imply, after some manipulations, that
$\bar{\Xi}_{v}^{ab}=-\tilde{\nabla}^{[a}\xi^{b]}$ (the so-called Kosmann lift~\cite{Kos66}), where
$\tilde{\nabla}$ is the covariant derivative with respect to the standard transposed connection on
$\Sig_{\rho}$. On the other hand, the Lie derivative of spinor fields can be expressed 
in terms of $\bar{\Xi}_{v}^{ab}$
as follows:
\beq
\pounds_{\bar\Xi}\psi= \xi^{a}e_{a}\psi +\frac{1}{4} \hat{\Xi}_{ab}\gam^{a}\gam^{b}\psi = 
\xi^{a}\nabla_{a}\psi-\frac{1}{4}\nabla_{[a}\xi_{b]}\gam^{a}\gam^{b}\psi=\\=
\xi^{\mu}\der_{\mu}\psi +\frac{1}{4}\hat{\Xi}_{h\,ab}\gam^{a}\gam^{b}\psi - \frac{1}{4}
\nabla_{[a}\xi_{b]}\gam^{a}\gam^{b}\psi\,,
\eeq
where $\hat{\Xi}_{h}$ is the horizontal part of $\hat{\Xi}$ {\it with respect to 
the spinor-connection}. 
We see that, {\it because of relation \eqref{oh!oh!}}, once the expression of 
$\bar{\Xi}_{v}^{ab}$ derived by Eq. \eqref{liederspin} has been substitued in Eq. \eqref{oh!},
we obtain a condition involving the spinor-connection $\tilde{\ome}$.

This result has been pointed out in Ref.~\cite{Winterroth07}. It agrees with an analogous 
one obtained in Ref.~\cite{FFR03} within a different
approach to conservation laws for the Einstein--Cartan theory of gravitation.
 
In Ref.~\cite{GoMa03,GoMa05} a  
geometric interpretation of the Kosmann lift as a reductive lift has been recovered
for the definition of a $\textstyle{SO}(1,3)^{e}$-reductive Lie derivative of spinor fields.
We justify the ``naturality'' of the Kosmann lift from a variational point of view: it charaterizes 
the only gauge-natural lift
which ensures the naturality condition 
$\cL_{j_{s+1}\hat{\Xi}_{H}}[\cL_{j_{s+1}\hat{\Xi}_{V}}\lam]\equiv 0$ holds true. Along
such a lift not only the initial Lagrangian $\lam$ is by assumption invariant, 
but also its first variational derivative, {\it taken with respect to vertical parts of gauge-natural lifts lying in the kernel of the generalized Jacobi morphism},
$\ome(\lam,\mathfrak{K})$ is it, thus implying that either $[\hat{\Xi}_{H},\hat{\Xi}_{V}]=0$ or 
it is a gauge-natural lift.
 The Hamiltonian content of such
a naturality condition and implications on conserved quantities have been investigated in
Refs.~\cite{FPW05,PaWi06,Winterroth07}.
The two approaches are strictly related: in a forthcoming paper~\cite{PaWi08} 
we study how the kernel of the Jacobi
morphism induces a split structure which is also reductive.  
 
\section*{Acknowledgments}

Research supported by MIUR--PRIN (2005) and University of Torino. 
Thanks  are due to Prof. I. Kol{\'a}{\v r} for interesting discussions and
for the kind invitation to take part to the DGA2007 Conference. 


\end{document}